\documentclass[superscriptaddress,twocolumn,floatfix]{revtex4}

\usepackage{graphicx,graphics,epsfig}   % Include figure files
\usepackage{amsmath,amsfonts,amssymb}    % need for subequations
\usepackage{amsthm,graphicx,color}

\usepackage{float}
\usepackage{hyperref}

\newcommand{\be}{\begin{equation}}
\newcommand{\ee}{\end{equation}}
\newcommand{\ba}{\begin{eqnarray}}
\newcommand{\ea}{\end{eqnarray}}
\newcommand{\ban}{\begin{eqnarray*}}
\newcommand{\ean}{\end{eqnarray*}}

\newcommand{\blk}{\color{black}}
\newcommand{\blu}{\color{blue}}
\definecolor{DarkGreen}{rgb}{0,.5,0}

\newcommand*{\tr}{\mathsf{Tr}}

\newcommand{\ket}[1]{\left|{#1}\right\rangle}

\begin{document}

%\preprint{APS/123-QED}

\title{Experimental semi-device-independent certification of entangled measurements}

\author{Adam Bennet}
\affiliation{Centre for Quantum Dynamics and Centre for Quantum Computation and
Communication Technology, Griffith University, Brisbane, 4111, Australia}
\author{Tam\'as V\'ertesi}
%\email{tvertesi@dtp.atomki.hu}
\affiliation{ Institute of Nuclear Research of the Hungarian Academy of Sciences
H-4001 Debrecen, P.O. Box 51, Hungary}
\blu
\author{Dylan J. Saunders}
\affiliation{Centre for Quantum Dynamics and Centre for Quantum Computation and
Communication Technology, Griffith University, Brisbane, 4111, Australia}
\affiliation{Clarendon Laboratory, University of Oxford, Parks Road, Oxford OX1 3PU, UK}
\author{Nicolas Brunner}%
\affiliation{D\'epartement de Physique Th\'eorique, Universit\'e de Gen\`eve, 1211 Gen\`eve, Switzerland}
\author{G. J. Pryde}
\affiliation{Centre for Quantum Dynamics and Centre for Quantum Computation and
Communication Technology, Griffith University, Brisbane, 4111, Australia}

\date{\today}

\begin{abstract}
Certifying the entanglement of quantum states with Bell inequalities allows one to guarantee the security of quantum information protocols independently of imperfections in the measuring devices. Here we present a similar procedure for witnessing \textit{entangled measurements}, which play a central role in many quantum information tasks. Our procedure is termed semi-device-independent, as it uses uncharacterized quantum preparations of fixed Hilbert space dimension. Using a photonic setup, we experimentally certify an entangled measurement using measurement statistics only. We also apply our techniques to certify unentangled but nevertheless inherently quantum measurements.
\end{abstract}

%\pacs{03.65.Ud, 03.67.Mn, 42.50.Dv}

\maketitle

%----------------------------------------------------------------------------------

\emph{Introduction.---} Entanglement is nowadays viewed as the paradigmatic feature of quantum theory. Entanglement underpins quantum information science, where it represents a powerful resource for information processing, secure communication and precision measurement. In recent decades entanglement has been demonstrated and carefully characterized in a wide range of physical platforms, with the strongest demonstrations employing Bell inequalities \cite{bell,review}, where violation of the inequality implies that the underlying quantum state is entangled. Importantly, this verdict is \textit{device-independent}, in the sense that it does not rely on any assumption about the alignment of the measurement devices, or of the Hilbert space dimension of the state. Device-independent entanglement verification is therefore of great practical importance, for example in systems where it is difficult to guarantee the precise configuration of measuring devices, due for instance to unnoticed side-channels. Moreover, the device-independent approach ensures security in realistic implementations of adversarial tasks, such as cryptography \cite{DI,prx}. 

\begin{figure}[b!]
\begin{center}
\includegraphics[width=0.95\linewidth]{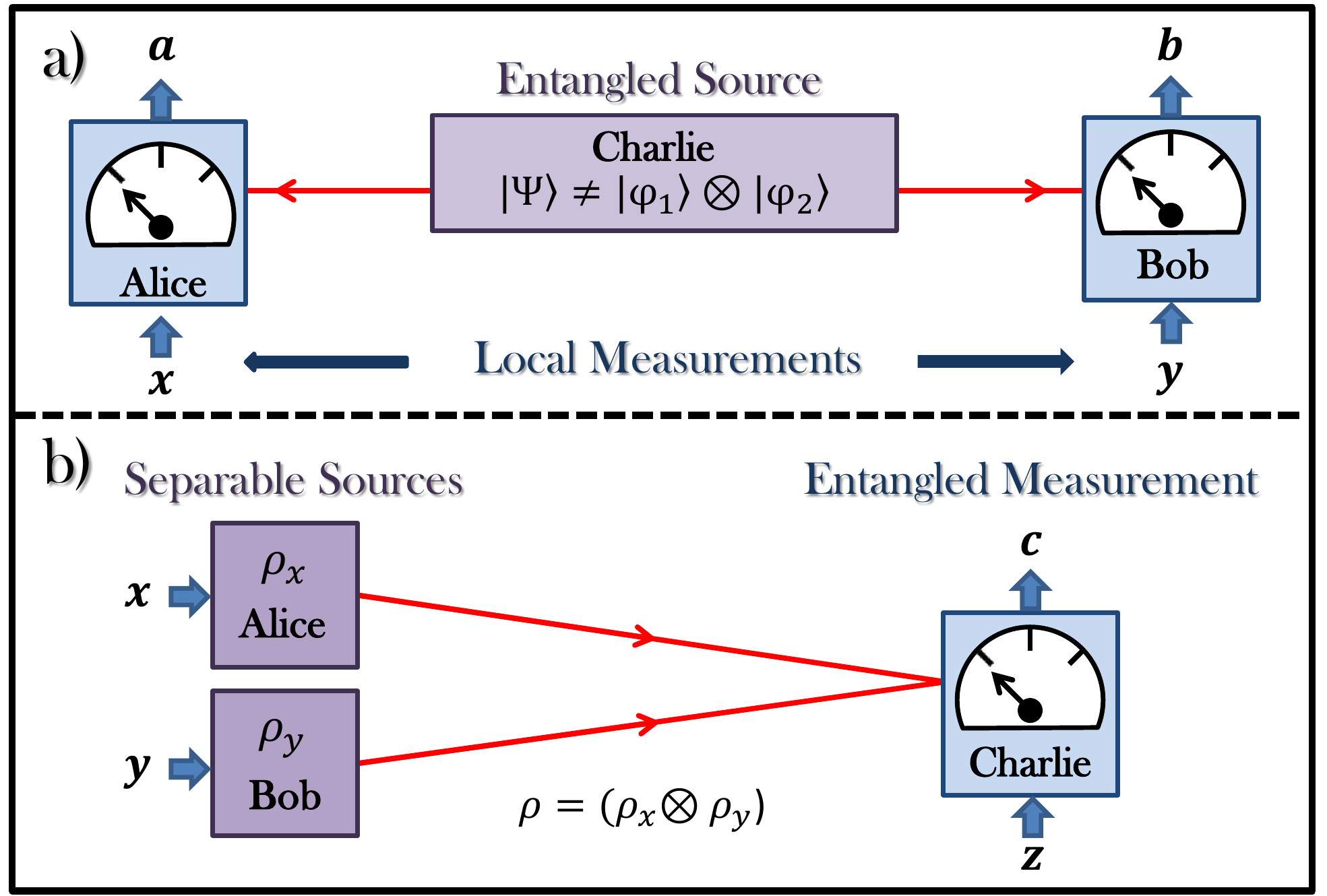}
\end{center}
\vspace{-5ex}
\caption{Conceptual representation of (a) a Bell inequality test, in which product measurements are performed on an entangled state, and (b) our scenario, in which product states are measured jointly, and projected onto an entangled state. In (a) the entanglement of the shared state $\ket{\Psi}$ can be certified from the observed data, i.e the probability distribution $p(a,b|x,y)$ (where $x,y$ denote measurement settings, and $a,b$ the corresponding outcomes), via Bell inequality violation. In (b), the presence of an entangled measurement can be certified from the data $p(c|x,y,z)$ using our witness for entangled measurements. Here we require the assumptions that the sources are uncorrelated, i.e. $\rho=\rho^A_{x}\otimes\rho^B_{y}$, and produce qubit states.} 
\label{fig:concept}
\end{figure}

Quantum mechanics also allows for \textit{entangled measurements}, a concept which complements the preparation of entangled states (Fig.~\ref{fig:concept}). Specifically, an entangled measurement is described by an operator for which at least one of the eigenstates corresponds to an entangled state. Entangled measurements have been studied much less than entangled states, however, such measurements play a fundamental role in many manifestations of quantum information science, including quantum teleportation \cite{tele}; dense coding \cite{DC}; parameter estimation \cite{para}; quantum repeaters \cite{sangouard}; and quantum computation \cite{NS}. Hence, the verification and characterization of entangled measurements constitutes a critically important task. 

One standard approach for measurement characterization is quantum detector tomography - essentially the analogue of quantum state tomography but for a measurement process \cite{luis,fiurasek,Lundeen09}. These approaches are restriced in generality because they require access to a well-calibrated source of quantum states spanning the state space, a feat not enforcable consistently. It is therefore important to ask whether this stringent state-preparation requirement can be dispensed with. Fundamentally, it is also valuable to know whether or not a measurement can be certified as entangled based solely on measurement statistics, in the same way that quantum states can be. In a recent theoretical work, V\'ertesi and Navascu\'es \cite{VN} showed such verification is indeed possible, under the assumption that the prepared states used to test the measuring device are of fixed Hilbert space dimension (but otherwise uncharacterized). This approach, termed \textit{semi-device-independent}, has been followed for other tasks, such as the quantification of entanglement \cite{liang}, cryptography protocols \cite{marcin} and randomness certification \cite{li}. Related works expound certification of the presence of an entangled measurement in a fully device-independent scenario \cite{rabello} (see also \cite{tzyh}), however, these approaches are of limited experimental interest as they are not robust to noise and/or involve experimentally unfeasible measurements.

Here we demonstrate how entangled measurements can be certified experimentally in the semi-device-independent framework. We present theoretically a simple and robust test of entangled measurements, and show that a partial linear optics Bell state measurement device \cite{braunstein95} conveniently produces the optimal quantum violation. We demonstrate this optimal strategy in a photonic experiment and test for violation of our witness. We also explore the possibility of discriminating between quantum \textit{un}entangled measurements and purely classical measurements. We construct a simple witness for this problem and implement it experimentally. Our work shows that semi-device-independent techniques are useful in experimental quantum information tasks, and thus complement recent experiments on device-independent estimation of quantum system preparations \cite{hendrych,ahrens,blatt}.

%----------------------------------------------------------------------------------

\emph{Scenario.---} We consider three separated parties, Alice, Bob, and Charlie. Alice and Bob each hold a preparation device, which we suppose emits uncharacterized qubit states. Each party can choose between $n$ possible preparations, labelled $x=0,...,n-1$ and $y=0,...,n-1$ respectively. The corresponding qubit states are denoted $\rho^A_x$ and $\rho^B_y$. Since Alice and Bob are separate and independent, the joint state that they prepare is unentangled. The states prepared by Alice and Bob are then transmitted to Charlie, who holds an uncharacterized measurement device. Upon receiving both input states, Charlie chooses a measurement setting $z$, and the device provides a measurement outcome $c$. Our scenario can thus be viewed as the dual (or time-reversed) version of a standard Bell test (see Fig.~\ref{fig:concept}).

Let us denote by $\{M_{c|z} \}$ the elements of Charlie's positive operator valued measure (POVM). After repeating this protocol many times, the parties obtain the probability distribution of each outcome $c$ given any possible pair of preparations $x,y$ and measurement $z$, i.e.
\begin{equation}
p(c|x,y,z) = \tr(\rho^A_x \otimes \rho^B_y \cdot M_{c|z}). \label{prob}
\end{equation}
This represents the experimental data. Our goal is to identify the type of measurement performed in Charlie's device, based only on the data. 
The key point is to distinguish between several classes of measurements: 

\begin{description}
  \item{\textbf{Classical measurements:}} Alice's and Bob's devices each send one (classical) bit of information, denoted $b_A$ and $b_B$, to Charlie's device, which outputs according to an arbitrary function $c=f(b_A,b_B,z)$.
  \item{\textbf{LOCC measurements:}} The measurement corresponds to a sequence of local measurements on Alice and Bob's individual qubits, such that each measurement possibly depends on the outcomes of earlier measurements.
  \item{\textbf{Unentangled measurements:}} Each of the POVM elements $M_{c|z}$ is a separable operator (for all $c,z$).
  \item{\textbf{Entangled measurements:}} At least one of the POVM elements $M_{c|z}$ is not separable.
\end{description}

Note that we have the inclusion relations: General meas. $\supset$ Unentangled meas. $\supset$ LOCC meas. $\supset$
Classical measurements, where General meas. refers to the set of all quantum measurements, including entangled and unentangled ones.

%----------------------------------------------------------------------------------

\emph{Robust test for entangled measurements.---} We present a simple test for certifying the presence of an entangled measurement. Consider a party, Charlie, who inherits a measurement box that takes two qubits as inputs and correspondingly yields one of three classical outputs. Charlie inquires into the claim that the measurement performed by the device is an entangled quantum measurement. A procedure for witnessing entangled measurement is to allow $n=3$ preparations for each of the two qubits, a task which will be assigned to Alice and Bob respectively. 
Charlie's device performs a fixed (single-setting) ternary measurement with outcome $c=1,2,3$ (hence the index $z$ is omitted). We consider the linear witness 
\begin{equation}
w=\sum_{c=1,2}\sum_{x,y=0}^2 {W_{c|x,y}p_{c|x,y}}, \label{w}
\end{equation}
where the coefficients are given by
\begin{equation}
W_{1|x,y}=\left(
\begin{tabular}{ccc}
 1 & -1 & -1\\
-1 &  1 & -1\\
-1 & -1 &  1\\
\end{tabular}
\right),\, W_{2|x,y}=
\left(
\begin{tabular}{ccc}
 1 & -1 & -1\\
-1 & -1 &  1\\
-1 &  1 & -1\\
\end{tabular}
\right). \label{coeff}
\end{equation}

Next we derive the maximal value for our witness \eqref{w} for the different classes of measurements discussed above. 
Since our witness is linear, it follows that the maximal value for a classical measurement strategy can be obtained for a deterministic strategy. As there is only a finite number of such strategies, we exhaustively evaluate them and find that $w \leq w_{classical}=1$. This bound can be obtained by considering the following strategy: Alice (Bob) sends $b_A=1$ ($b_B=1$) iff $x=0$ ($y=0$), and Charlie outputs $c=1$ upon receiving $b_A=b_B=1$, and $c=3$ otherwise. 

For unentangled measurements we find that $w\leq w_{unent}= 1$. This represents our witness for detecting entangled measurements: measurement statistics producing $w>1$ cannot be obtained from any strategy using unentangled measurements. Note that the bound $w_{unent}= 1$ was obtained via numerical methods (see-saw iteration \cite{pal}); however, the modest complexity of the problem and the large number of iterations provides very strong evidence of optimality. From the inclusion relations mentioned above we also get that $w_{LOCC}\leq 1$ because $w_{general}\geq w_{ent} \geq w_{unent} \geq w_{LOCC} \geq w_{classical}$.

Next, we observe that entangled measurement can outperform unentangled measurements, in the sense of giving a larger value of $w$. Consider Alice and Bob's preparations to be the pure qubit states represented by Bloch vectors
$\vec r_x =(\cos\alpha_x,0,\sin\alpha_x)$ and $\vec q_y
=(\cos\beta_y,0,\sin\beta_y)$, where $\rho^A_x = (\openone + \vec
r_x\cdot\vec \sigma)/2$, $\rho^B_y = (\openone + \vec q_y\cdot\vec
\sigma)/2$ and $\vec\sigma=(\sigma_x,\sigma_y,\sigma_z)$ is the vector of Pauli matrices. For Charlie's device, we consider the POVM elements
\ba
M_1 =|\phi^+\rangle\langle \phi^+| \, , \,\, M_2 =|\phi^-\rangle\langle \phi^-| \, , \,\, M_3 =\openone - M_1 - M_2, \label{partial}
\ea
where $\ket{\phi^\pm}=(\ket{00}\pm \ket{11})/\sqrt{2} $. This POVM represents the so-called partial Bell state measurement, a routinely used and deterministically implementable measurement in linear optics \cite{Mattle96}. Overall we obtain 
\begin{align}
p(c|x,y) =\tr{(\rho^A_x\otimes\rho^B_y M_c)}=\frac{1}{4}(1+\cos(\alpha_x + (-1)^c \beta_y))
\end{align}
for $c=1,2$  (the $c=3$ condition is found by normalisation but plays no role in the witness). Setting $\alpha_{j} =\beta_{j}=2\pi j/3$ for $j=0,1,2$ we obtain $w=3/2$, hence largely exceeding the bound for unentangled measurements. 
In fact, we verified using numerical methods (see-saw) that the above strategy is optimal, i.e. $w_{ent}=3/2$, representing the maximal possible value allowed by quantum mechanics (considering qubit preparations).

%----------------------------------------------------------------------------------

\emph{Robust test of unentangled versus classical measurements.---} We now address the question of discriminating unentangled quantum measurements from classical measurements, again for \blk the case of $n=3$ preparations. However, Charlie now chooses between two dichotomic measurements, $z=0,1$, with outcome $c=1,2$. We consider the linear witness
\begin{equation}
v=\sum_{z=0,1}\sum_{x,y=0}^2 {V_{c=1|x,y,z}p_{c=1|x,y,z}}. \label{w2}
\end{equation}

For simplicity, we omit the notation $c=1$ and just write $V_{x,y,z}$ and $p_{x,y,z}$. The coefficients of the witness are given by
\begin{equation}
V_{x,y,z=0}=\left(
\begin{tabular}{ccc}
 2 &  0 &  0\\
 0 & -2 & -2\\
 0 & -2 & -2\\
\end{tabular}
\right),\, V_{x,y,z=1}=
\left(
\begin{tabular}{ccc}
 0 &  0 &  0\\
 0 &  1 & -1\\
 0 & -1 &  1\\
\end{tabular}
\right). \label{coeff2}
\end{equation}
As above, the maximal value for a classical measurement can be obtained by checking all deterministic strategies. We find $v_{class}\leq 2$, which is our witness for detecting unentangled but non-classical measurements. Note that the bound $v_{class}=2$ can be obtained by considering the following strategy: Alice (Bob) sends $b_A=1$ ($b_B=1$) iff $x=0$ ($y=0$), and Charlie outputs $c=1$ upon receiving $b_A=b_B=1$, and $c=2$ otherwise. 

The following strategy demonstrates that unentangled measurements can outperform classical measurements. Consider as above Alice and Bob preparing qubit states in the $x,z$ plane of the Bloch sphere, given by the angles $\alpha_{j} =\beta_{j}=2\pi j/3$ for $j=0,1,2$. Charlie's POVM elements (for outcome $c=1$) are given by
\begin{align}
M_{1|z=0} &=|0\rangle\langle 0|\otimes |0\rangle\langle 0|,\nonumber\\
M_{1|z=1} &=|+\rangle\langle +|\otimes |+\rangle\langle +|+|-\rangle\langle -|\otimes |-\rangle\langle -|,
\label{unent}
\end{align}
where $|\pm\rangle = (|0\rangle\pm|1\rangle)/\sqrt{2}$. Overall, we obtain $v=3$ which clearly outperforms classical measurements. 

%----------------------------------------------------------------------------------

\emph{Experimental verification of entangled measurements.---} The tests presented aim to reveal the nature of an unknown measurement based solely on the measurement statistics of a device. We experimentally certify entangled measurements, as well as unentangled (but nevertheless genuinely quantum) measurements via our witnesses, using photonic polarization qubits, linear quantum optics, and single photon counting modules (SPCMs).

\begin{figure}[t!]
\begin{center}
\includegraphics[width=0.95\linewidth]{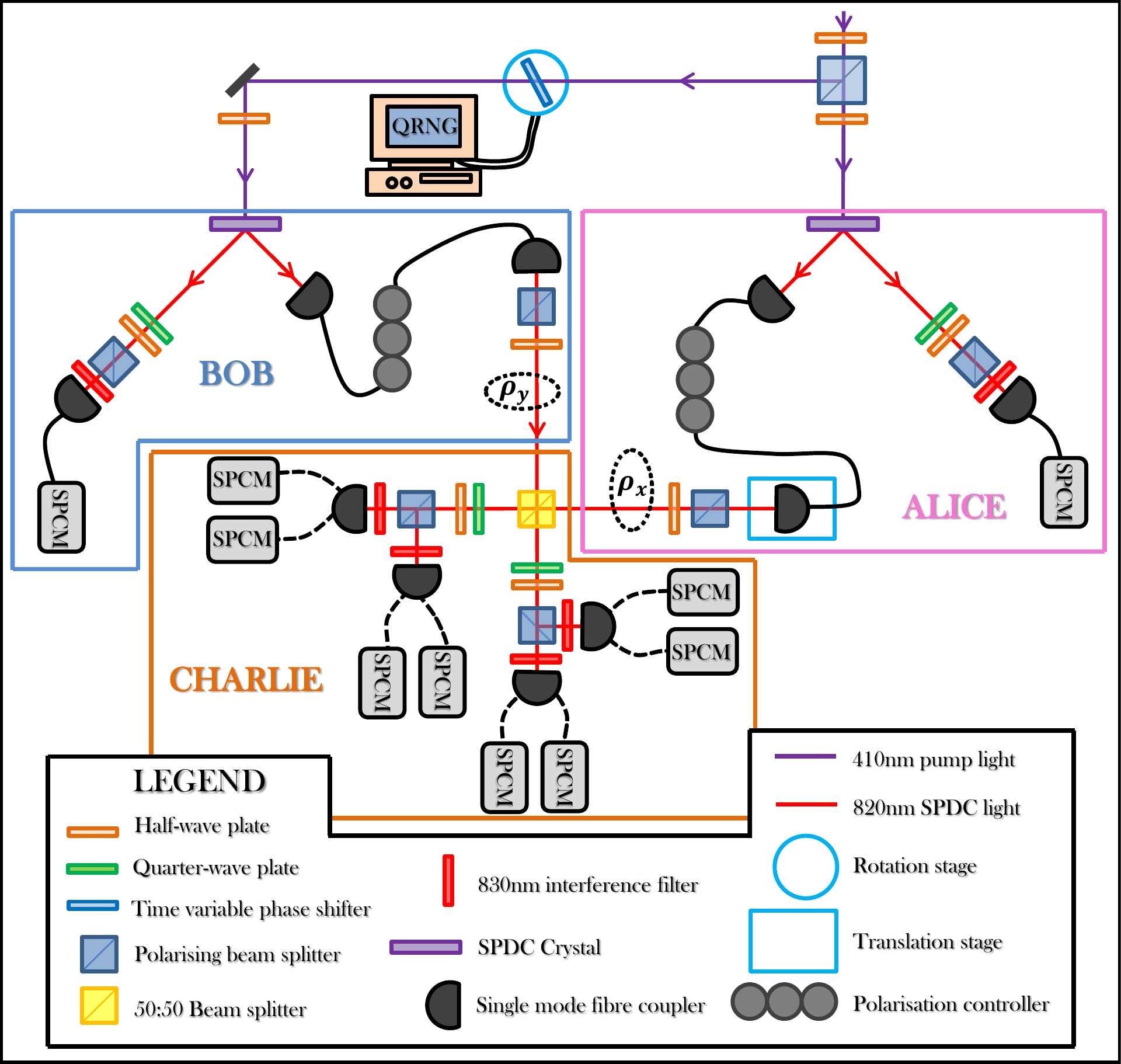}
\end{center}
\vspace{-5ex}
\caption{Experimental apparatus. A pair of separate spontaneous parametric down conversion (SPDC) sources create Alice's and Bob's photon pairs. One photon from each pair of Alice's and Bob's sources act as heralding signals, with the remaining photons sent via optical fibre (solid lines) to the inputs of Charlie's partial BSM device. The type of measurement Charlie enacts is changed with the translation stage. Dashed lines represent outputs from a 50:50 fibre beam splitter. The time variable phase shifter (TVPS) is a glass plate on a rotation stage connected to an online quantum random number generator (QRNG)  \cite{anu}, its purpose being to erase coherence between the pump beam shared by Alice and Bob, enforcing source independence and separability.}
\label{fig:Exp}
\end{figure}

The entangled measurement was realised via a partial Bell-state measurement (BSM) device based on a beam splitter and polarization analysis \cite{Mattle96}. Depending on photodetector ``click'' patterns, this device determines projections onto the singlet state $\ket{\Psi^-}=\left(\ket{HV}-\ket{VH}\right)/\sqrt{2}$, the triplet state $\ket{\Psi^+}=\left(\ket{HV}+\ket{VH}\right)/\sqrt{2}$, or the remaining triplet subspace spanned by $\{\ket{\Phi^{\pm}}\}=\{ \left(\ket{HH}\pm \ket{VV}\right)/\sqrt{2} \}$, where $H$ and $V$ denote horizontal and vertical polarizations. This is equivalent to the POVM elements $M_{1,2,3}$ of equation \eqref{partial}. The standard mode of operation for the BSM sees the device combine pairs of identical photons (meaning degenerate in every degree of freedom, except polarization where information is encoded) on a 50:50 beam splitter. The photons undergo non-classical Hong-Ou-Mandel (HOM) interference \cite{HOM}, succeeded by polarization analysis and two photon pseudo-number-resolving detection (pseudo number resolution is achieved using 50:50 fiber beam splitters and SPCMs at the output couplers - Fig. \ref{fig:Exp}).

This same device was used to enact quantum \textit{un}entangled measurements by making conditions for non-classical two-photon interference unfavourable. We achieved this by delaying the relative arrival time between the photon pairs incident on the 50:50 beam splitter, enforcing temporal distinguishability. To test the action of Charlie's measurement (which in principle is unknown), Alice and Bob sent pairs of initially separable photons into Charlie's measurement device. To create separable input states Alice and Bob individually employed type-I spontaneous parametric down-conversion sources (using BiBO crystals) to generate polarization-unentangled pairs of single photons. Because Alice and Bob's photons were generated in independent down-conversion sources, their polarization encoded qubits were also unentangled in all other degrees of freedom, justifying the restricted dimensionality assumption of this semi-device-independent technique \cite{note2} (more information on the photon source is provided in the Appendix). The remaining photons from each of Alice's and Bob's pairs acted as heralding signals for the entanglement verification protocol. Before being received by Charlie, Alice and Bob's photons were polarization encoded (using motorized half-wave plates) in each of the $n=3$ preparations in the $x$, $z$ plane of the Bloch sphere, creating states $\rho^{A}_{x}$ and $\rho^{B}_{y}$. 

\begin{table}[h!]
\vspace{10pt}
\begin{center}
\begin{tabular}{|c|c|c|c|}
\hline
Meas't witnessed & Bound & Experiment & Upper limit \\
\hline
\hline
Entangled & $w=1$ & $w_{exp}=1.32\pm0.07$ & $w=1.5$ \\
\hline
Unent. quantum & $v=2$ & $v_{exp}=2.75\pm0.06$ & $v=3$ \\
\hline
\end{tabular}
\end{center}
\caption{Inequality violations for quantum entangled and quantum unentangled measurements. For the former, $1<w_{exp}<1.5$ implies entanglement. For the latter, $2<v_{exp}<3$ implies nonclassical measurement. Uncertainties in $w_{exp}$ and $v_{exp}$ are derived from single photon counting statistics.}
\label{Table:Data}
\end{table}

We first investigated the case of an entangled measurement, when Charlie's device performs a partial BSM. The observed statistics resulted in $w_{\textrm exp}$ = 1.32$\pm$0.07, violating the bound for unentangled measurements ($w_{\textrm unent}=1$) by more than four standard deviations. Hence from the statistics of the experiment only, we can guarantee that Charlie's device performs an entangled measurement. Note that we did not reach the maximal possible violation of the witness, i.e. $w_{ent}=3/2$, primarily due to imperfect HOM visibility. Simulations using our imperfect HOM visibility ($\sim90\%$) lowered the expected value of the witness to $w \sim 1.37$, consistent with our data.  

Next, we moved to the case of an unentangled measurement, when Charlie's device performs the two possible dichotomic measurements given in equation \eqref{unent}. From the statistics of the experiment, we evaluated our second witness (see equation \eqref{w2}) and obtained $v_{exp}$ = 2.75$\pm$0.06, hence violating the bound for purely classical measurements ($v_{class}=2$) by more than 12 standard deviations. As above, imperfect HOM visibility accounts for the reduced violation of the witness (theoretically we had $v_{unent}$ = 3). 

The experimental data processing was subjected to standard assumptions for an estimation scenario. First, we assumed that the observers were free to choose the preparations and measurements. Next, we assumed independent trials, that is, in each run of the experiment the statistics are described by equation (1). Finally, we assumed the observed statistics represented a fair sample of the total statistics, which would be obtained with detectors having unit efficiency. 
%----------------------------------------------------------------------------------

\emph{Conclusion.---} 
Entangled measurements and quantum operations are ubiquitous to many modern quantum information protocols. The distribution of entanglement in a quantum network through entanglement swapping, and many other applied and fundamental quantum tests, require the certification and characterization of entangled measurements. Using the semi-device-independent approach, we have presented and experimentally implemented a simple and efficient witness for verifying the presence of (i) entangled measurements, and (ii) unentangled but nevertheless inherently quantum measurements. As our tests are based on measurement statistics only, they provide a very practical and powerful tool for the estimation of quantum systems, and should find application in many quantum information protocols. Our technique can in principle also be used to certify an unknown process as entangling (such as a controlled-NOT gate), since such processes can implement an entangled measurement \cite{note1}.

%----------------------------------------------------------------------------------

\emph{Ackowledgements.} This research was partially supported by the Australian Research Council (ARC) Centre of Excellence for Quantum Computation and Communication Technology (project number CE110001027). G.J.P acknowledges and ARC Future Fellowship. N.B. acknowledges financial support from the Swiss National Science Foundation (grant PP00P2\_138917), the EU DIQIP and COST.

\bibliographystyle{prsty}

\begin{thebibliography}{10}
\bibitem{note2} Using a beam splitter, we deliver frequency-doubled pump laser light from the same laser to both SPDC crystals. However, we randomize the phase of one of the pump beams, after the beam splitter, to ensure that the pump modes are unentangled, further guaranteeing the independence of the photon generation events.
\bibitem{note1} For example, a controlled-NOT can be configured to implement an entangled measurement \cite{Langford05}. The semi-device independent approach provides a certification that requires less calibration than the more complex quantum process tomography \cite{Rohde05}.
\end{thebibliography}

%----------------------------------------------------------------------------------

\begin{widetext} 
\newpage

\appendix
%\section{Appendix}
\section{Experimental details}

To demonstrate Charlie's entangling measurement witnesses, we constructed two independent sources of photon pairs. The two sources comprised optically nonlinear Bismuth Borate crystals (BiBO; BiB$_{3}$O$_{6}$), each type-I phase matched for spontaneous parametric downconversion from 410nm (pump, horizontally polarised) to 820nm (signal/idler, degenerate, vertically polarised). One photon from each pair of Alice's and Bob's sources acted as heralding signals, with the remaining photons encoded in polarisation states ($\rho_{x}$ and $\rho_{y}$ respectively) and sent via optical fibre to the inputs of Charlie's partial BSM device. Optical bandpass filters (3nm FWHM, Semrock) were placed at each output coupler connected to single photon counting modules (SPCMs, Perkin Elmer SPCM-AQR-14-FC and custom single photon counting arrays). Coincidence events are registered using a field programable gate array with a coincidence window of 3ns. \\

%----------------------------------------------------------------------------------

\noindent The states prepared by Alice and Bob are described by Bloch vectors $\vec r_x =(\cos\alpha_x,0,\sin\alpha_x)$ and $\vec q_y=(\cos\beta_y,0,\sin\beta_y)$ where $\alpha_{j} =\beta_{j}=2\pi j/3$ for $j=0,1,2$, with corresponding polarisation state vectors $|A_{j}\rangle$ and $|B_{j}\rangle$. Importantly, for matters of experimental convenience, our partial BSM device was constructed to resolve the $|\Psi^{+}\rangle$ and $|\Psi^{-}\rangle$ Bell states, lumping the remaining $|\Phi^{\pm}\rangle$ outcomes into a third ``undecided'' outcome. This meant that, in order to optimally test the witnesses presented (which were designed to optimally resolve $|\Phi^{+}\rangle$/ $|\Phi^{-}\rangle$), the measurement required a $\sigma_{x}$ rotation on one qubit. Thus, in our experiment, Alice prepared the three original polarisation states $|A_{j}\rangle$ for each $j=0,1,2$, being $|H\rangle$ and $\frac{1}{2}|H\rangle\pm\frac{\sqrt{3}}{2}|V\rangle$, and Bob prepared the three original polarisation states with a $\sigma_{x}$ rotation applied, those being $|V\rangle$ and $\frac{1}{2}|V\rangle\pm\frac{\sqrt{3}}{2}|H\rangle$. \\

%----------------------------------------------------------------------------------

\noindent To ensure accurate measurement and qubit preparation, all preparation and measurement wave plates were in computer controlled motorised rotation states (Newport).To characterise Alice's and Bob's preparations, we measured the expectation values for the projection operators in the H and V polarisation basis ($\hat{\Pi}_{H}$ and $\hat{\Pi}_{V}$ respectively) for each of Alice's and Bob's preparations and compared them with theory, shown in Table \ref{table:Proj} below.  \\

\begin{table}[h!]
\begin{center}
\begin{tabular}{c|c|c|c|c}
 \hline
 State ($|A_{j}\rangle$,$|B_{j}\rangle$) & $\langle\hat{\Pi}_{H}^{Theory}\rangle$ & $\langle\hat{\Pi}_{V}^{Theory}\rangle$ & $\langle\hat{\Pi}_{H}^{Exp}\rangle$ & $\langle\hat{\Pi}_{V}^{Exp}\rangle$ \\
\hline
\hline
$|A_{0}\rangle=|H\rangle$ & 1 & 0 & 0.993 & 0.007 \\
\hline
$|A_{1}\rangle=\frac{1}{2}|H\rangle+\frac{\sqrt{3}}{2}|V\rangle$ & 0.25 & 0.75 & 0.256 & 0.744 \\
\hline
$|A_{2}\rangle=\frac{1}{2}|H\rangle-\frac{\sqrt{3}}{2}|V\rangle$ & 0.25 & 0.75 & 0.251& 0.749 \\
\hline
\hline
$|B_{0}\rangle=|V\rangle$ & 0 & 1 & 0.009 & 0.991\\
\hline
$|B_{1}\rangle=\frac{1}{2}|V\rangle+\frac{\sqrt{3}}{2}|H\rangle$ & 0.75 & 0.25 & 0.750 & 0.250 \\
\hline
$|B_{2}|\rangle=\frac{1}{2}|V\rangle-\frac{\sqrt{3}}{2}|H\rangle$ & 0.75 & 0.25 & 0.743 & 0.257 \\
\end{tabular}
\end{center}
\caption{State preparation for Alice and Bob. Uncertainties in experimental quantities are less than $\pm$0.001.}
\label{table:Proj}
\end{table}

\noindent The boundary between Charlie's measurements being entangling measurements or unentangled quantum measurements was selected by enforcing or removing non-classical two-photon interference, set by the relative arrival time between the photon pairs incident on the 50:50 beam splitter in Charlie's device. The relative arrival times were dependent on the coherence length of the photons ($\approx$300$\mu m$), so separating their arrival by $>300\mu m$ eliminated non-classical two-photon interference, controlled by moving a motorized translation stage in Alice's arm (Fig.~2 of the main manuscript). Charlie's partial BSM device resolved the $\ket{\Psi^{+}}$ and $\ket{\Psi^{-}}$ Bell states through discrimination of orthogonally polarised photon pairs on the measurement beam splitters (the case of $\ket{\Psi^{+}}$), or non-classical antibunching behaviour arising from the antisymmetric singlet state (the case of $\ket{\Psi^{-}}$). The $\ket{\Phi^{\pm}}$ states, on the other hand, required number resolving detection (since these states saw pairs of photons degenerate in polarisation bunched at the point of detection). Because our SPCMs cannot implement number resolving operations, we instead opted for pseudo-number resolution, by replacing single fibres at Charlie's measurement with 50:50 fibre beam splitters. In theory, the pairs of photons encountering these splitters were resolved 50$\%$ of the time. In practice, we scaled the probabilities of the $\{\ket{\Phi^{\pm}}\}$ outcomes in accordance with the measured splitting ratios. \\

%----------------------------------------------------------------------------------

\noindent Alice and Bob's photons were generated by independent, random SPDC events in separate crystals and thus were unentangled in any degree of freedom, even though both crystals are pumped with the same laser. To further strengthen the assertion of independence, we introduced into the experiment a time variable phase shifter (TVPS) that erased coherence between the pump beams of Alice's and Bob's sources. This device was a thick glass plate on an angular rotation stage connected to a remote quantum random number generator \cite{anu} that imparted a random phase between $0$ and $2\pi$ on Bob's pump beam every 502ms. Over the course of the experiment, the phase shifter erased any entanglement between the two pump modes which might be thought to be converted into entanglement between the downconverters. 

%----------------------------------------------------------------------------------

\end{widetext}

\end{document}